          [2001/04/25 1.1 (PWD)]
\documentclass[letterpaper,twocolumn]{esapub} 

\usepackage{natbib, graphicx}

\title{Methanol abundance in low mass protostars}
\author{S\'ebastien Maret}
\affil{Department of Astronomy, University of Michigan, 833 Dennison
  Building, 501 East University Avenue, Ann Arbor MI 48109-1090, USA}

\newcommand{\Xout}{X_\mathrm{out}}
\newcommand{\Xin}{X_\mathrm{in}}

\def\hydrogen{H$_2$\ }

\def\formaldehyde{H$_2$CO\ }
\def\methanol{CH$_3$OH\ }

\defcitealias{Maret04}{Paper~I}

\begin{document}


\maketitle

\begin{abstract}

Methanol lines observations of a sample of low mass Class 0 protostars
are presented. Using a 1D radiative transfer model, I show that
several protostars have large abundance jumps in the inner hot and
dense region of envelopes, probably because of thermal grain mantle
evaporation. These abundances are compared with a grain surface
chemistry model.

\end{abstract}

\section{Introduction}

During the formation of a star, the gas undergoes important changes,
both on a physical and chemical point of view. In the prestellar
phase, the gas is heavily depleted on grain mantles. When the
gravitational collapse starts, the protostar gradually warm up the gas,
and the trapped molecules are released in the gas phase. These
molecules can then rapidly trigger the formation of more complex
molecules, by gas phase chemistry. Methanol is a well suited molecule
to study these changes, because it is abundant in grain mantles, and
is expected to evaporate from grain mantles in the inner hot and dense
regions of the protostars. Methanol observations can therefore be used
to determine the physical and chemical conditions in these regions. In
this contribution, I present methanol lines observations a sample of
low mass protostars. Using a detailed radiative transfer model, I
derive the methanol abundances in the envelopes, and discuss the
implications on the formation of this molecule.

\section{Observations and results}

\begin{figure}
  \centering
  \includegraphics[width=\columnwidth]{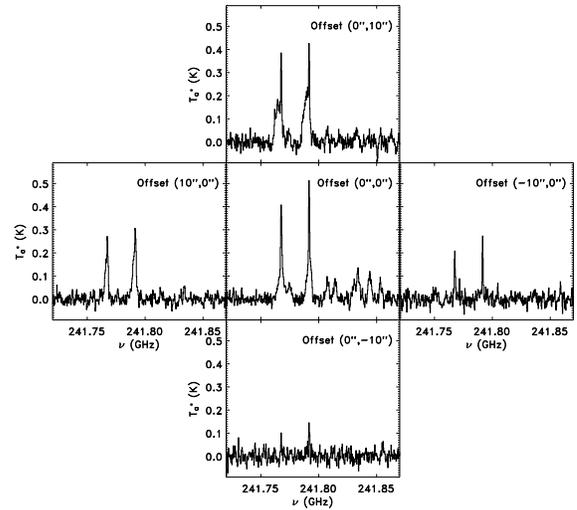}
  \caption{Methanol 5$_K$-4$_K$ maps of NGC1333-IRAS2.}
  \label{fig:map_iras2}
\end{figure}

Six Class 0 protostars \citep{Andre00} were observed. The methanol
5$_K$-4$_K$ lines were observed with the IRAM-30m telescope, while the
7$_K$-6$_K$ were observed with the JCMT. Fig. \ref{fig:map_iras2} show
an example of methanol 5$_K$-4$_K$ map obtained with the IRAM-30m
on NGC1333-IRAS2. Lines have gaussian profile with high velocity
wings. Narrow high energy lines ($>$ 100 K) are detected on the
central position. These lines likely originate from the inner and
dense part of the envelope, while broader, low energy lines clearly
originate from the two outflows detected in CO 3-2 and 2-1
\citep{Knee00}, SiO 2-1 \citep{Blake96,Jorgensen04b} and CS 3-2
\citep{Langer96}.

\section{Model}

\begin{figure}
  \centering
  \includegraphics[width=\columnwidth]{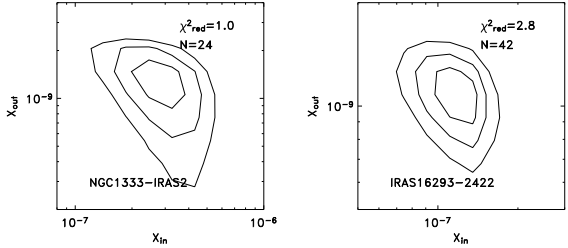}
  \caption{$\chi^2$ as a function of the inner and outer methanol
    abundances for NGC1333-IRAS2 and IRAS16293-2422. The contours
    indicates the 1, 2, and 3 $\sigma$ confidence levels
    respectively. The number of observed lines, $N$, are also shown, as
    well as the reduced $\chi^2$ for the best fit model.}
  \label{fig:chi2}
\end{figure}

In order to determine the methanol abundances in the envelopes, a
detailed 1D radiative transfer model has been developed. The model
uses the escape probability formalism, and is presented into details
in \citet[][hereinafter Paper I]{Maret04}. \methanol collisional rates
with para-\hydrogen from \citet{Pottage04} are used. The envelopes are
assumed to be spherically symmetric, and the density profile is
supposed to follow a power law. Index of the power law and dust
temperature profile from \citet{Jorgensen02} and \citetalias{Maret04}
are used. The methanol abundance profile is supposed to follow a step
function. The abundance is set to constant value $\Xout$, in the outer
cold part of the envelope. This abundance jumps to a $\Xin$ value in
the inner and hotter part of the envelope, because of the evaporation
of the grains mantles. This evaporation is assumed to occurs at a
temperature of 100 K. The best fit values $\Xin$ and $\Xout$ are
determined by minimizing the $\chi^2$ between the model and the
observations. Fig. \ref{fig:chi2} show the $\chi^2$ maps obtained for
NGC1333-IRAS2 and IRAS16293-2422.

\section{Results and discussion}

The methanol abundances in the outer part of the envelope range from
$3 \times 10^{-10}$ to $2 \times 10^{-9}$, and are quite similar from
one source to the other. On the contrary, the inner abundance varies
much from one source to the other. In two sources of our sample,
NGC1333-IRAS2 and IRAS16293-2422, the observations can only be
reproduced by our model if there are jumps in the abundances. The
inner abundances are $2 \times 10^{-7}$ and $1 \times 10^{-7}$
respectively, i.e. a factor 100 and 200 larger than in the outer
envelope. In NGC1333-IRAS4B and L1448-MM, there are weak evidences of
abundances jumps (1 $\sigma$). In these sources, the inner abundance
jump to $7 \times 10^{-7}$ and $5 \times 10^{-7}$ respectively, i.e. a
factor about 300 larger than in the outer cold part of the
envelope. These values should however taken with caution, because of
the small confidence levels. In NGC1333-IRAS4A, L1448-N and L1157-MM,
the observations are well reproduced with a constant CH$_3$OH
abundance throughout the envelope, even if the presence of a jump can
not be ruled out by the present observations.

Gas phase chemistry reactions can not reproduce the high abundances of
methanol observed in ices (Herbst, private communication). Methanol is
therefore likely to be be formed on grain surfaces by successive
hydrogenation of CO \citep{Tielens82}. Because methanol is evaporated
from grain mantles in the inner part of the envelopes of our sample,
these abundances are likely to reflect the composition of ices, and
can be use to determine the efficiency of the formation of this
molecule on grains mantles.I have compared the abundances obtained in
the four protostars of our sample where abundance jumps were detected
with a grain chemistry model \citep{Keane04}. Both CO
\citep{Jorgensen02}, \formaldehyde \citepalias{Maret04} and \methanol
(this study) were found to be well reproduced by the model. The model
indicates that density at the time of the formation of the molecule is
$10^{-5}\ \mathrm{cm}^{-3}$. It also indicates that the probability of
H to react with CO and H$_2$CO are equal, in agreement with recent
laboratory experiment \citep{Hidaka04}.


\section{Conclusion}

The present observations and model show that methanol abundance is greatly
enhanced in the inner regions of protostellar envelope, probably
because thermal evaporation of grain mantles. The observed abundances
are well explained by a grain surface chemistry model, and indicates
that methanol was formed at a density of $\sim$ 10$^{-5}$ cm$^{-3}$,
probably during the prestellar phase.

\section*{Acknowledgments}

I wish to thank Eric Herbst for a useful discussion about the methanol
formation.

\bibliographystyle{aa}
\bibliography{marets}

\end{document}